\documentclass{osa-article}

\journal{oe}

\articletype{Research Article}

\begin{document}

\title{Inverse Design of Invisibility Cloaks using the Optical Theorem}

\author{Brian Slovick\authormark{*} and Josh Hellhake}

\affil{Advanced Technology and Systems Division, SRI International, 333 Ravenswood Avenue, Menlo Park CA, 94025, USA}

\email{\authormark{*}brian.slovick@sri.com} 

\begin{abstract}
We develop and apply an optimization method to design invisibility cloaks. Our method is based on minimizing the forward scattering amplitude of the cloaked object, which by the optical theorem, is equivalent to the total cross section. The use of the optical theorem circumvents the need to evaluate and integrate the scattering amplitude over angle at each iteration, and thus provides a simpler, more computationally efficient objective function for optimizing structures. We implement the approach using gradient descent optimization and present several gradient-permittivity cloaks that reduce scatter by metallic targets of different size and shape.
\end{abstract}

\section{Introduction}

The development of invisibility cloaks is a longstanding goal in electromagnetics \cite{Devaney2012,Colton2012,Fiddy2014,Wolf1993,Rumpf2007,Hoenders1997,Gbur2003}. An object that is invisible to electromagnetic waves would have widespread application in both military and commercial systems. At radio frequencies, cloaked objects have zero radar cross section \cite{Knott2004,Jenn2005}, while in the optical band, displays with low scattering and reflection provide an improved viewing experience \cite{Raut2011,Chattopadhyay2010}. It has also been proposed that cloaked sensors may provide enhanced sensitivity owing to a reduced interaction with the environment \cite{Alu2009,Bilotti2011,Greenleaf2011,Chen2011,Fan2012}.

More recently, the development of metamaterials and metasurfaces has expanded the design space to include materials with negative refractive index and plasmonic properties \cite{Zheludev2012}. Materials with negative index can be used with transformation optics to design electromagnetic cloaks \cite{Xi2007,Chen2010,Pendry2012}, and several implementations have been demonstrated \cite{Schurig2006,Valentine2009,Kante2009}. Transformation optics has the considerable advantage of imposing no limitations on the size and shape of the object. However, the designs often require complex anisotropic materials with negative refractive index, which inevitably suffer from large absorption loss and dispersion \cite{Khurgin2015}.

Scattering cancellation is an alternative method for achieving invisibility \cite{Chen2012}. Of course, antireflection coatings are used througout industry to reduce reflection by planar surfaces \cite{Raut2011,Chattopadhyay2010}. For nonplanar structures, such as spheres and cylinders, multilayered designs involving dielectric \cite{Qiu2009,Huang2007} and plasmonic materials \cite{Edwards2009,Silveirinha2008} have been proposed. However, these designs are either limited to simple shapes or suffer from large absorption losses \cite{Khurgin2015}. Patterned metasurfaces, or mantle cloaks, have been proposed as an alternative to thicker, multilayer designs \cite{Alu2009b,Liu2014}. However, their application is limited to subwavelength objects. Therefore, a more generalized design approach applicable to objects of any size and shape, that does not involve plasmonics or materials with negative index, would be a significant advance.

In recent years, inverse methods have become the predominant approach in photonic design \cite{Tahersima2019,Su2018,Jin2018,Hughes2018,Liu2018,Slovick2020,Sitawarin2018,Molesky2018}. Broadly described, inverse design is the optimization of an objective function with respect to the structure or material properties. Inverse design has been used to optimize photonic circuits \cite{Tahersima2019,Su2018,Jin2018} and nanophotonic resonant structures \cite{Hughes2018,Liu2018,Slovick2020,Sitawarin2018,Molesky2018}. The key to developing an effective inverse design algorithm lies in the definition of the objective function. For instance, to design highly scattering structures, a suitable objective function is the determinant of the wave operator defining the poles of the scattering matrix \cite{Slovick2020}. On the other hand, nonscattering structures can be designed by minimizing the scattering cross section \cite{Bondeson2004,Chaudhury2009}. However, the need to calculate the cross section from the scattered field for all angles at each step in the optimization poses a considerable computational challenge, and limits the scale and type of objects that can be optimized.

In this Letter, we develop a more efficient optimization method for designing invisibility cloaks. Our inverse design approach is based on minimizing the forward scattering amplitude, which by the optical theorem is equal to the total cross section \cite{Boya1994,Hovakimian2005}. The use of the optical theorem circumvents the need to integrate the scattered fields over angle by evaluating the cross section from the forward scattering amplitude alone, and thus provides a simpler, more computationally efficient algorithm. Our method is completely general and makes no assumptions about the size, shape, or composition of the object, though large objects still present a computational challenge. For demonstration, we apply it to design gradient-permittivity cloaks to minimize scatter from metallic targets of different size and shape. We show that orders of magnitude reductions in cross section are achievable with subwavelength coatings. The ease and effectiveness of our approach enables the optimization of large-scale nonscattering structures and invisibility cloaks.

\section{Approach}
Consider an object in free space with relative permittivity $\epsilon_r(\textbf{r})$. In the scalar approximation of electromagnetics, the field scattered by the object is obtained by solving the Helmholtz equation with radiation boundary conditions. The solution for the scattered field is \cite{Devaney2012,Colton2012,Fiddy2014}
\begin{equation}
E_s(\textbf{r})=\int G(\textbf{r}-\textbf{r}')F(\textbf{r}')E(\textbf{r}')d^3r',
\end{equation}
where $G(\textbf{r}-\textbf{r}')$ is the Green's function given by 
\begin{equation}
 G(\textbf{r}-\textbf{r}')= \frac{e^{ik|\textbf{r}-\textbf{r}'|}}{|\textbf{r}-\textbf{r}'|},
 \end{equation}
where $F(\textbf{r})=k^2[\epsilon_r(\textbf{r})-1]/(4\pi)$ and $E(\textbf{r})$ is the total field given by the sum of the scattered field and the incident field $E_i(\textbf{r})$, which we assume is a plane wave of the form $e^{i\textbf{k}_i\cdot \textbf{r}}$. A similar expression can be obtained for the full electromagnetic vector field \cite{Liu2019,Slovick2020}, but here we only consider the scalar form. In the far field approximation, the Green's function reduces to
\begin{equation}
 G(\textbf{r}-\textbf{r}')\simeq \frac{e^{ikr}}{r}e^{-i\textbf{k}\cdot \textbf{r}'},
 \end{equation}
where $r=|\textbf{r}|$ and $\textbf{k}=k\textbf{r}/r$ is the scattered wavevector. The scattered far field is then
 \begin{equation}
E_s(\textbf{r})=\frac{e^{ikr}}{r}f(\textbf{k},\textbf{k}_i),
\end{equation}
where $f(\textbf{k},\textbf{k}_i)$ is the scattering amplitude given by  \cite{Devaney2012,Colton2012,Fiddy2014}
\begin{equation}
    f(\textbf{k},\textbf{k}_i)= \int e^{-i\textbf{k}\cdot \textbf{r}'} F(\textbf{r}')E(\textbf{r}')d^3r'.
\end{equation}
At this stage, the scattering cross section is normally evaluated by integrating the square modulus of the scattering amplitude over solid angle. This expression can be used as the objective function to design non-scattering structures \cite{Bondeson2004,Chaudhury2009}. However, the need to evaluate and integrate the scattering amplitude over angle leads to a significant computational expense. A much simpler method is to apply the optical theorem \cite{Boya1994,Hovakimian2005}, which states that the total cross section is proportional to the imaginary part of the scattering amplitude evaluated in the forward ($\textbf{k}=\textbf{k}_i$) direction as
\begin{equation}
    \sigma=\frac{4\pi}{k} \text{Im} [f(\textbf{k}_i,\textbf{k}_i)].
\end{equation}
The optical theorem implies that the scattering amplitude evaluated in just one direction can be used as the objective for a minimization function. However, to evaluate the scattering amplitude and cross section using Eqs. (5) and (6), we need an expression for the total field. This can be accomplished by discretizing space as $\textbf{r}(j)=jh$, where $h$ is the grid size. In discrete form, Eq. (1) can be written as \cite{Ying2015,Liu2019}
\begin{equation}
E_s(jh)=\sum_i h^3 G(jh-ih)F(ih)E(ih),
\end{equation}
or in matrix-vector form
\begin{equation}
\textbf{E}_s=GF \textbf{E},
\end{equation}
where $F$ is a diagonal matrix and $h^3$ has been absorbed into the definition of $G$. This matrix equation can be solved to obtain the total field as \cite{Ying2015,Liu2019,Slovick2020}
\begin{equation}
\textbf{E}=(I-GF)^{-1}\textbf{E}_i.
\end{equation}
Substituting this expression into the discrete form of Eq. (5), we obtain
\begin{equation}
    \sigma=\frac{4\pi}{k} \text{Im} \left[ \textbf{E}^*_iF(I-GF)^{-1}\textbf{E}_i \right].
\end{equation}
 Equation (10) forms the objective function of our inverse design optimization algorithm. For a particular incident direction, we optimize nonscattering structures by minimizing $\sigma$ with respect to the permittivity values using a nonlinear least squares algorithm. Note, that in the continuum limit $h\rightarrow 0$, Eq. (10) is an exact expression for the scattering amplitude, accounting for all multiple scattering and resonance effects.
 
 The most computationally intensive calculation in Eq. (10) is the matrix inversion, and the desire is to minimize the number of matrix inversions required to minimize the scattering cross-section. Conventional gradient descent algorithms would need to evaluate the derivative of the objective with respect to $F(\textbf{r})$, which would involve a numerical evaluation of the Jacobian and an additional matrix inversion. Fortunately, since the objective function has an analytical form, we can obtain a closed-form expression for the Jacobian as
 \begin{equation}
     \frac{\delta}{\delta F} F(I-GF)^{-1}\textbf{E}_i=\left[I+F(I-GF)^{-1} G\right](I-GF)^{-1}\textbf{E}_i.
 \end{equation}
This expression provides a rapid determination of the Jacobian matrix at each step in the gradient descent, and more importantly, only requires the same matrix inversion already performed to calculate the cross-section. This removes extraneous matrix inversions at each step in the gradient descent and greatly speeds up the nonlinear least squares algorithm.

We tested this methodology using Matlab on a standard PC, and the memory limitations provided a practical limitation on the size matrix that could be inverted, thus limiting the size of the scattering object compared to the wavelength. To relax this size limitation, we tested the methodology using a 2D scattering object. Using a 2D structure required rederiving Eq. (1) with an integral over two dimensions and the Green’s function for the 2D Helmholtz operator \cite{Fiddy2014,Ying2015}
\begin{equation}
    G_{2D}(\textbf{r}-\textbf{r}')=\frac{i}{4}H_0^{(1)}(k|\textbf{r}-\textbf{r}'|),
\end{equation}
where $H_0^{(1)}(x)$ is a Hankel function of the first kind. While this form of the Green’s function is used in the transfer matrix, the far field form of Eq. (4) used in the scattered field equation is
\begin{equation}
        G_{2D}(\textbf{r}-\textbf{r}')\simeq \sqrt{\frac{i}{8\pi k r}}e^{ikr}e^{-i\textbf{k}\cdot \textbf{r}'}
\end{equation}
Repeating the optical theorem derivation for the 2D case results in the same expression for the scattering cross-section as Eq. (10), with the volume elements replaced by area elements \cite{Boya1994,Hovakimian2005}.

\section{Results}
To test this methodology for designing nonscattering objects, we considered various cases of an optimized cloak applied to a metal rod. The model was parameterized, with dimensions of the metal rod, the cloak, and the grid element dimensions all defined relative to a wavelength $\lambda$. The permittivity of the metal rod was chosen to be $\epsilon_{rod} =-120 + 4i$, based on the measured value for copper at 1.5 $\mu$m \cite{Babar2015}.  However, we quickly found that neither the resulting cloak design or the scattering cross-section reduction were dependent on the copper permittivity for values ranging from microwave to infrared frequencies. Figure 1 shows the model setup, which in this case included the solid copper cylinder with 0.5$\lambda$ radius, a 0.5$\lambda$ thick cloak, and grid size of 0.0625$\lambda$, resulting in 812 grid elements across the rod and cloak. The plane wave was incident in the $+\hat{\textbf{x}}$ direction with a polarization parallel to the length of the rod. The top row of plots  (Fig. 1a) shows the permittivity distribution with no cloak as well as the scattered and total field magnitudes inside and outside the rod calculated using the discrete integral form of Eqs. (8) and (9).  As expected, we see that the rod strongly scatters the incident wave creating interference patterns throughout the region around the rod.

\begin{figure}
\includegraphics[scale=0.32]{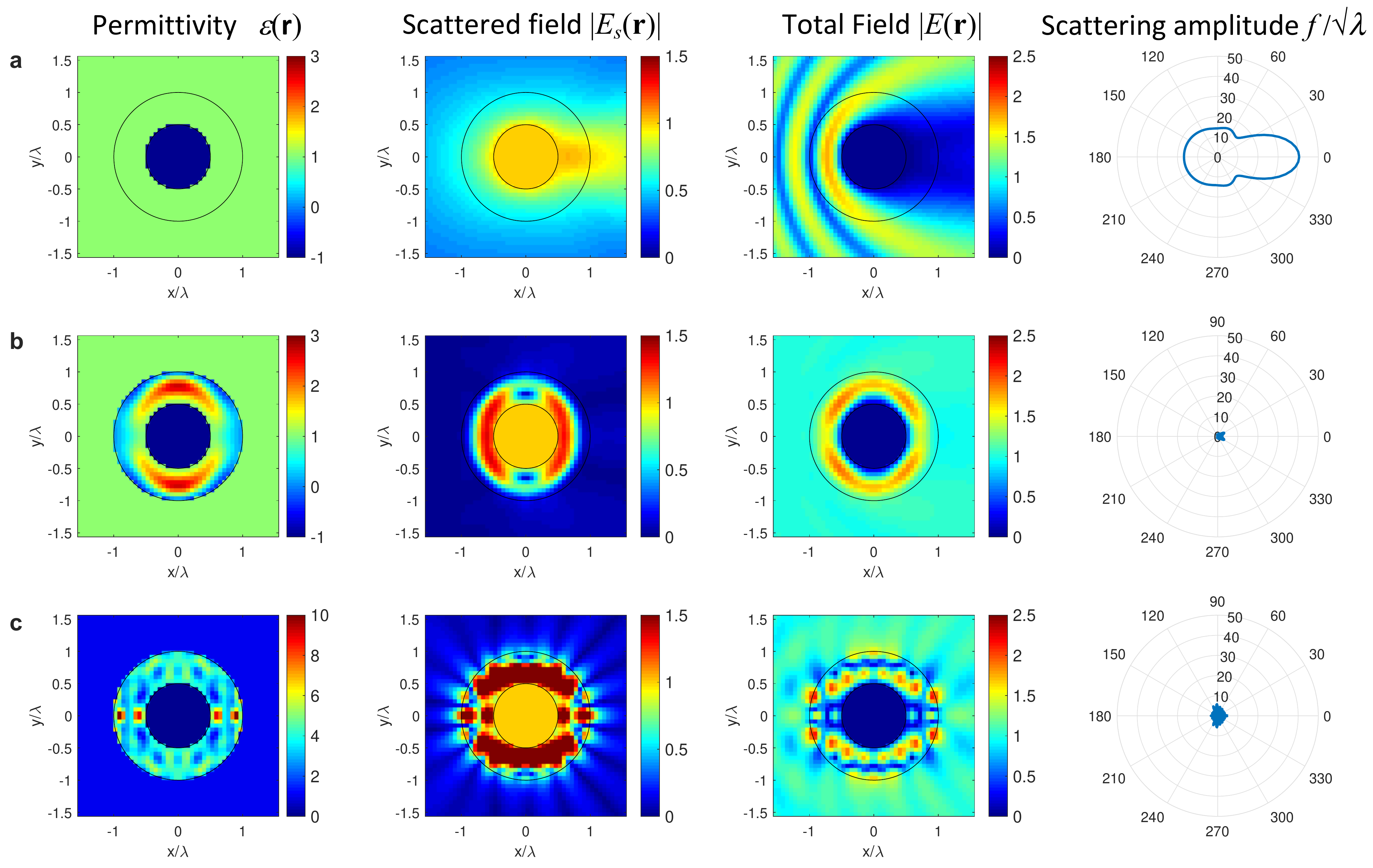}
\caption{Relative permittivity, scattered fields, total fields, and scattering amplitude for a 0.5$\lambda$ radius copper rod with (a) no cloak, (b) optimized 0.5$\lambda$-thick cloak for an initial seed of $\epsilon_{shell}=1$, and (c) optimized 0.5$\lambda$-thick cloak for an initial seed of $\epsilon_{shell}=4$.}
\end{figure} 

The second row (Fig. 1b) shows results of a non-linear least squares algorithm computed using the lsqnonlin function in Matlab. As non-linear least squares algorithms require an initial seed value for the optimization, we defined the initial shell permittivity to be 1 throughout the cloak. The result was a cloak permittivity distribution of all real values ranging from -1 to +3, roughly centered around the initial seed value of +1, and with an expected mirror symmetry which was not imposed in the Matlab algorithm. This design shows extremely small scattered field magnitudes around the cloak, resulting in a total field magnitude of near unity. Calculating the scattering cross section with and without the cloak using Eq. (10) shows a reduction factor of $\sim$1900. This strong reduction was confirmed by calculating the scattering amplitude, shown in the final column of Fig. 1b. These results confirm the effectiveness of this methodology and were generated in less than 1 minute of computational time.

\begin{figure}
\includegraphics[scale=0.32]{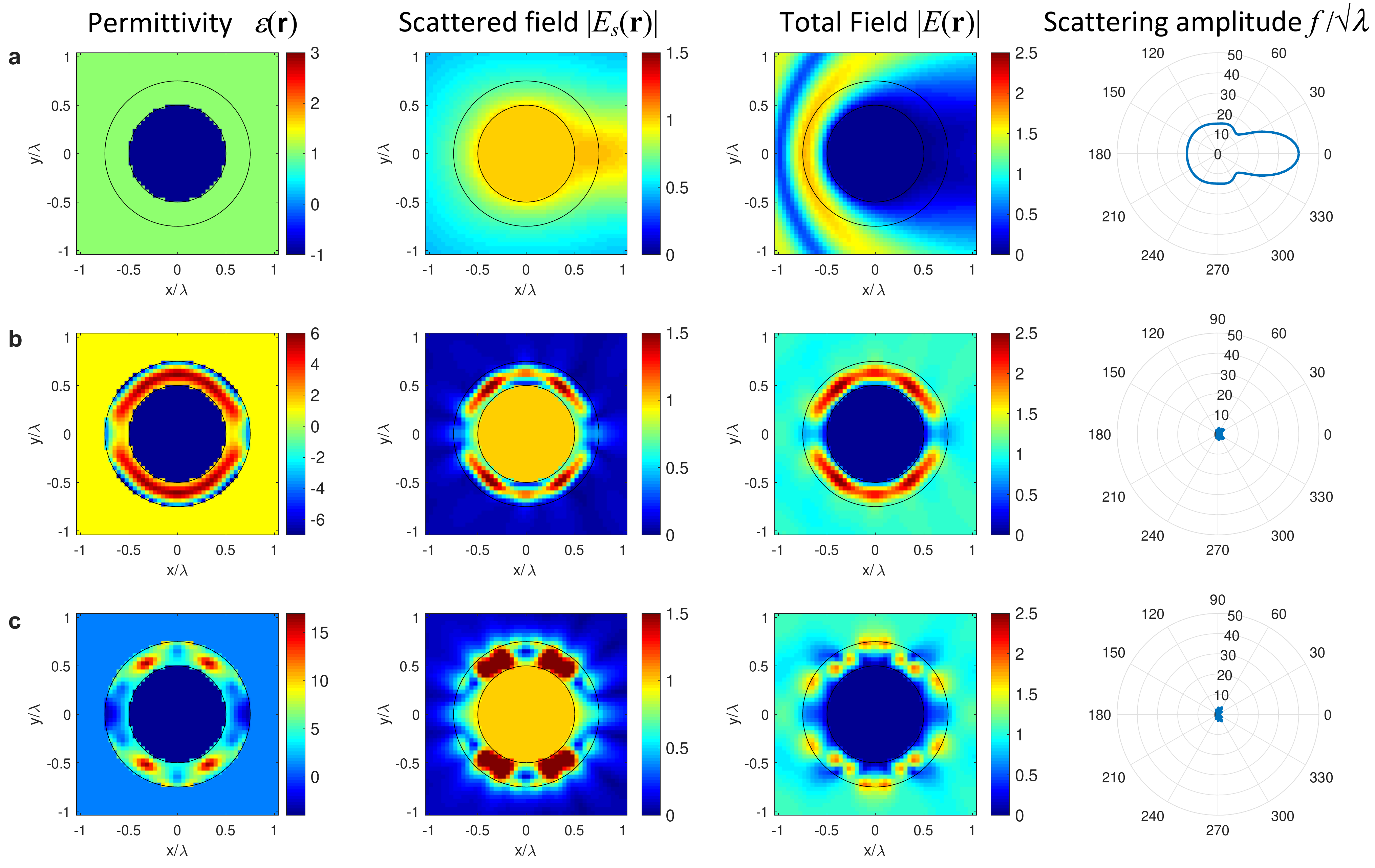}
\caption{Relative permittivity, scattered fields, total fields, and scattering amplitude for a 0.5$\lambda$ radius copper rod with (a) no cloak, (b) optimized 0.25$\lambda$-thick cloak for an initial seed of $\epsilon_{shell}=1$, and (c) optimized 0.25$\lambda$-thick cloak for an initial seed of $\epsilon_{shell}=4$.}
\end{figure} 

One limitation of the lsqnonlin function in Matlab is that it does not allow for bounds on the permittivity values when solving underdetermined problems such as Eq. (10). As a result, we ended up with permittivity values in the cloak that were real and less than 1, which is unrealistic. One way around this was to rerun the optimization with an initial seed permittivity of 4 throughout the cloak. The results of this run are shown in the bottom row (Fig. 1c). The resulting permittivity values generally trended higher, though in some cases still dropped below 1, and the permittivity distribution was more complex as compared to the simpler gradient for the prior seed permittivity. The resulting scatter reduction was still very good, with a reduction factor of $\sim$500, but clearly shows more structure in the near field scattering.

Having generated these promising results for a relatively thick cloak, we then performed similar optimizations with half the cloak thickness, 0.25$\lambda$. With the smaller structure, the grid size was also reduced to 0.0417$\lambda$.  The overall results in scattering reduction shown in Fig. 2 were still very large, with reduction factors of $\sim$340 and $\sim$150 for the $\epsilon_{shell}$=1 and $\epsilon_{shell}$=4 seeds, respectively.  However, the thinner cloak increased the range of permittivity values, resulting in permittivity values further below unity than in the previous design.

Lastly, we performed an additional cloak design for a metal rod with an elliptical cross-section in order to show applicability to general shapes aside from an ideal cylinder. In this case, the metal rod was an ellipse with a 1.2$\lambda$ major axis length and 0.4$\lambda$ minor axis length, with the major axis parallel to the incident field. The cloak was 0.2$\lambda$ thick around the elliptical rod, and the grid size was 0.025$\lambda$. The optimization results in this case, shown in Figs. 3b and c, gave cross-section reduction factors of $\sim$30 for both the $\epsilon_{shell}$=1 and $\epsilon_{shell}$=4 seeds. While not as high as the previous cases, these may be sufficient reductions in cross-section for many applications. Interestingly, for the $\epsilon_{shell}$=1 seed, the permittivity distribution looks qualitatively similar to the permittivity distribution for the cylinder in Fig. 2b, but distorted around the elliptical shape and with different extrema in the range of permittivity values. However, the scattered fields look quite different within the cloak.

\begin{figure}
\includegraphics[scale=0.32]{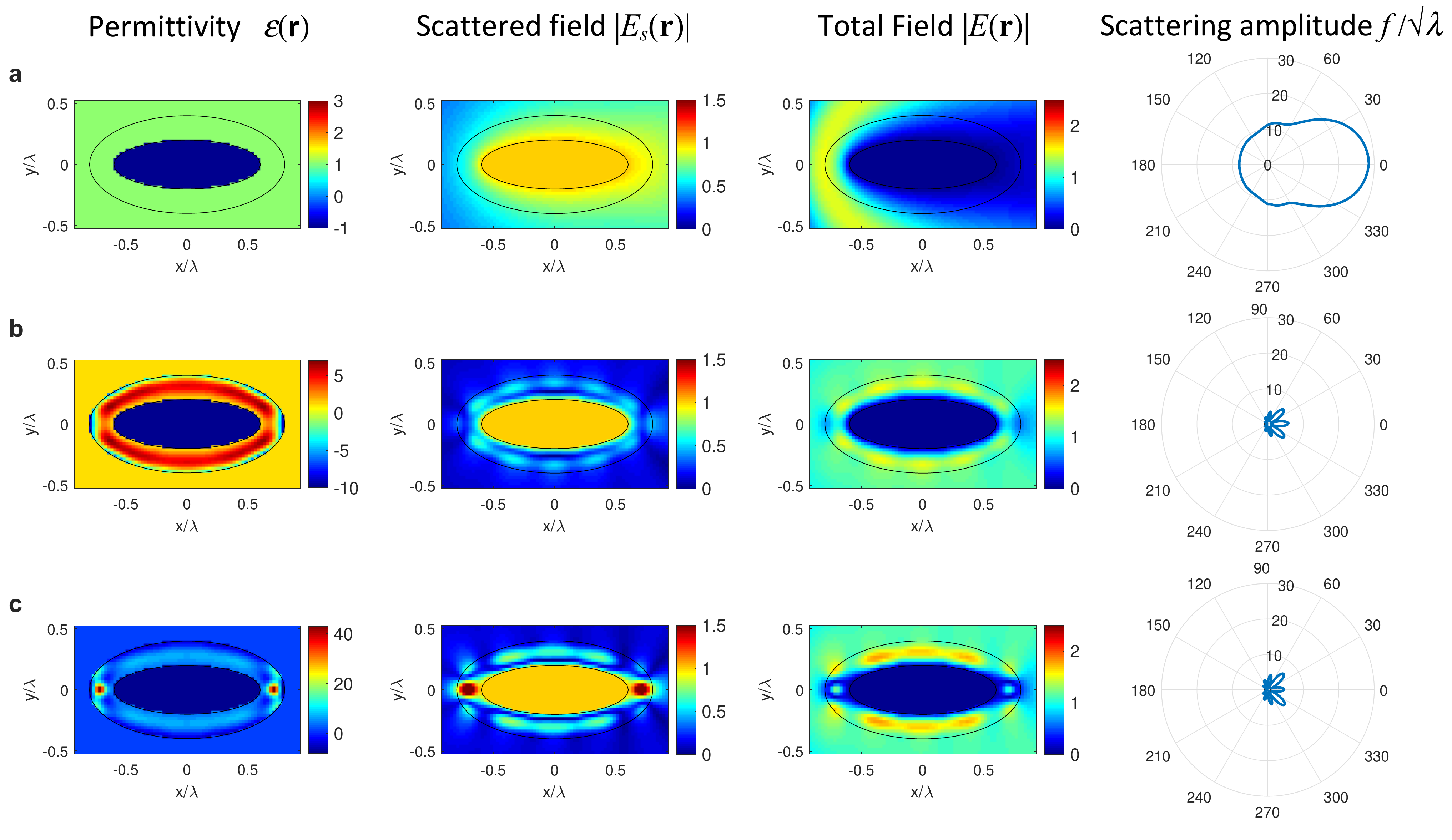}
\caption{Relative permittivity, scattered fields, total fields, and scattering amplitude for an elliptical copper rod with 1.2$\lambda$ and 0.4$\lambda$ major and minor axes, respectively, with (a) no cloak, (b) optimized 0.2$\lambda$-thick cloak for an initial seed of $\epsilon_{shell}=1$, and (c) optimized 0.2$\lambda$-thick cloak for an initial seed of $\epsilon_{shell}=4$.}
\end{figure} 

Overall, these cases all show that the optical theorem methodology for designing invisibility cloaks is highly effective. In all cases examined here, the optimization of the cloak resulted in cross-sections a small fraction of the initial metal rod, and the optimization was completed in minutes or less using a standard PC. Areas for future work include using a different or modified non-linear least squares algorithm which can employ bounds on the permittivity values, exploring the dependence of optimization seed values, and exploring the dependence of increasing grid density.

\section{Summary}
We developed and applied an inverse optimization method to design invisibility cloaks. The method is based on minimizing an objective function equal to the forward scattering amplitude of the cloaked object, which by the optical theorem, is equivalent to the total cross section. The use of the optical theorem greatly simplifies the optimization since only the forward scattering amplitude must be calculated at each iteration, in contrast to the traditional approach based on integrating the amplitude over angle. Using a nonlinear least squares gradient descent, we applied the method to design several gradient-permittivity cloaks to reduce scattering from metallic circular and elliptical cylinders. Our generalized approach provides a simple and effective design tool and enables the optimization of large-scale nonscattering structures and invisibility cloaks.

\section*{Funding}
Defense Advanced Research Projects Agency (DARPA) (HR001118C0015).

\section*{Disclosures}
The authors declare no conflicts of interest.

\bibliography{bib}

\end{document}